\documentclass[twocolumn,showpacs,preprintnumbers,amssymb,aps,pra,]{revtex4}

\usepackage{amsmath}
\usepackage{graphicx}
\usepackage{dcolumn}
\usepackage{bm}

\begin{document}

\title{Boson Pairs in a One-dimensional Split Trap}

\author{D.S.~Murphy}%
\author{J.F.~McCann}%
\email{j.f.mccann@qub.ac.uk} \affiliation{ Department of Applied
  Mathematics and Theoretical Physics, Queen's University Belfast,
  Belfast, BT7 1NN, Northern Ireland, UK}%
\author{J.~Goold}%
\author{Th.~Busch}%
\email{thbusch@phys.ucc.ie}\affiliation{ Department of Physics,
  National University of Ireland, UCC, Cork, Republic of Ireland}%

\date{\today}

\begin{abstract}
  We describe the properties of a pair of ultracold bosonic atoms in a
  one-dimensional harmonic trapping potential with a tunable
  zero-ranged barrier at the trap centre. The full characterisation of
  the ground state is done by calculating the reduced single-particle
  density, the momentum distribution and the two-particle
  entanglement. We derive several analytical expressions in the limit
  of infinite repulsion (Tonks-Girardeau limit) and extend the
  treatment to finite interparticle interactions by numerical solution.
  As pair interactions in double wells form a fundamental building 
  block for many-body systems in periodic potentials, our results have
  implications for a wide range of problems.
\end{abstract}


%
% 32.80.Pj - Optical cooling of atoms; trapping.
% 03.67.Mn - Entanglement production, characterization, and manipulation.
% 03.65.Ge - Solutions of wave equations: bound states.
% 05.30.Jp - Boson systems.
%

\maketitle

\section{Introduction}

The last two decades have seen considerable experimental advancement
in the area of cooling and trapping neutral atoms \cite{coh98}, with
one of the crowning achievements being the realization of Bose-Einstein
condensation (BEC) \cite{dal99}.  There continues, today, intense
experimental investigation into systems of trapped, ultracold atoms,
with potential deployment of this technology in the fields of
precision interferometry and quantum information processing
\cite{mon02}.  In particular, exciting advancements have been reported
on the behaviour of cold atoms in periodic potentials, which can be
created from the optical dipole forces arising from several crossed,
interfering laser beams \cite{bloA04,bloA05}.  Such arrangements have
allowed experimentalists to trap and control small numbers of
particles on tightly-confined, individual lattice sites and thereby
severely restrict their centre-of-mass dynamics.  Since such
potentials can be applied in selective directions in space, these
techniques allow the creation of effectively lower-dimensional systems
\cite{koh05,gor01,par04,kin04}.

A further external handle for control over cold atomic many particle
systems is the ability to change the interparticle scattering length,
allowing access to ideal, as well as strongly correlated regimes. This
can be accomplished by using Feshbach resonances \cite{zwi05} or by
tuning of the effective mass of particles moving in a periodic
potential \cite{par04}.

Combining these techniques has permitted the experimental realization
of atomic gases in the so-called Tonks-Girardeau (TG) regime, wherein
a quasi-1D quantum gas of strongly-interacting bosons acquires
fermionic properties \cite{ton36,gir60,yuk05}.  Not surprisingly,
these experimental advancements have motivated much theoretical
investigation of systems of strongly interacting bosonic gases in 1D,
subject to different confining potentials
\cite{gir00,gir01,bus03,lin07}.
 
The theoretical description of a sufficiently dilute system can be
achieved by restricting consideration to one- and two-particle effects, 
only.  As such the fundamental building block for the description of 
the many-body system is the system of two interacting particles, 
subject to some trapping potential. In addition, for the low momenta 
associated with ultracold particles it becomes possible to represent 
the particle-particle interactions through a pseudopotential, whereby 
the description of the particle-particle interactions depends only on 
the $s$-wave scattering length \cite{hua_st}.  Previous work has 
reported the analytical solution for a pair of particles in isotropic 
\cite{bus98} and anisotropic \cite{idz05} three-dimensional harmonic 
confining potentials, within the pseudopotential approximation.  The 
analytic solution for the 1D case is also presented in \cite{bus98}.  
It is straightforward to adapt this solution to the problem of a 
single particle in a $\delta$-split harmonic trap \cite{bus03}.  The 
delta-split trap potential may be viewed as a generic model for double
well situations or, alternatively, as a good approximation to the 
problem of a trap with an impurity at the centre.  In \cite{bus03} 
analytic 1D single-particle eigenstates are used to construct the 
many-body ground state for a system of $N$ particles confined by a 
$\delta$-split trap in the TG limit.

As the single-particle eigenstates are known for arbitrary barrier
strength, it is straightforward to obtain an analytic expression for the
two-particle ground state in the TG limit, while for finite
interactions a numerical scheme is required. In this work we analyse
the physics of a boson pair including the reduced single-particle
density, the momentum distributions and the two-particle entanglement,
which we quantify by means of the von Neumann entropy \cite{wan05, 
law05, sun06, pas01, li01, ghi03, ghiA04, ghiB04}.  In particular we
consider how these properties of the ground state may be altered as
both the barrier strength and interaction strength are varied.
Ultra-cold few-boson systems in a double-well trap have recently
received a thorough numerical investigation.  In \cite{zolA06, zolB06, 
zol07} the authors employ narrow-width Gaussians to model both the
contact potential and central splitting potential. They proceed to use
a Hartree-Fock type method to investigate some of the many-body
properties. The results of our work agree well with this numerical
approach in limiting cases.

The remainder of this paper is organized as follows.  In
Sec.~\ref{sect:model_hamiltonian} we describe the effective
Hamiltonian and the assumptions of the model.  In
Sec.~\ref{sect:single_particle} the single-particle eigenstates are
reviewed and used in Sec.~\ref{sect:two_bosons_tg} to construct an
analytical representation for the two-particle ground state in the TG
regime (i.e. the Tonks molecule).  Using this analytical 
representation we investigate the dependence on the barrier strength 
of the reduced single-particle density, the momentum distribution and
the von Neumann entropy for a boson pair.  
Sec.~\ref{sect:two_particles_numerical} employs a discretization 
scheme to allow for the variation of the interaction strength between
the particles. The computational method is outlined and a set of 
results are presented.  Finally, in Sec.~\ref{sect:conclusions} we 
make some concluding remarks and comment on the experimental 
realization of the proposed system.
 
\section{Model Hamiltonian}
\label{sect:model_hamiltonian}

Consider a system of two identical bosonic atoms which are confined in
a highly anisotropic harmonic trapping potential where the trapping
frequency in the perpendicular directions, $\omega_\perp$, is much
larger than in the axial direction, $\omega_{\perp}\gg\omega$. The
associated length scales are $d_\perp=\sqrt{\hbar/m\omega_\perp}$ and
$d = \sqrt{\hbar/m\omega}$.  As a result of the large energy level
separation, associated with the transverse eigenstates ($\hbar
\omega_\perp$), at low temperatures the transverse motion is 
restricted to the lowest mode.  In this case the system can be treated
as quasi-1D and may be described using the effective Hamiltonian,
\begin{equation}
 \label{eq:two_part_ham}
 \mathcal{H} = \sum_{i = 1,2} h_{i} + g_{1D} \delta( x_2 - x_1 )\;,
\end{equation}
where the single-particle Hamiltonian, $h_{i}$ is given by
\begin{equation}
  \label{eq:SingleParticleHamiltonian}
  h_{i}=-\frac{\hbar^2}{2m}\frac{\partial^2}{\partial x_{i}^{2}} 
        +\frac{1}{2}m\omega^{2} x_{i}^{2} + \kappa \delta (x_{i})\;.
\end{equation}
Here, $m$ is the particle mass, and $x_{i}$ ($i = 1, 2$) is the 1D 
position coordinate of particle $i$.  The last term of the 
single-particle Hamiltonian represents a point-like barrier located at
the origin and the parameter $\kappa>0$ determines the strength of 
this barrier.  The quantity $g_{1D}$ represents the interaction 
strength, and is related to the 1D s-wave scattering length ($a_{1D}$)
through $g_{1D} = -2\hbar^2/m a_{1D}\;$.  In turn, $a_{1D}$ is related
to the actual three-dimensional s-wave scattering length, $a_{3D}$, 
through $a_{1D} = -d_{\perp}^{2}/2a_{3D}(1 - Ca_{3D}/d_{\perp})\;$, 
where $C$ is a constant of value $C = 1.4603 \ldots$ \cite{ols98}.

In the limit of tight confinement, the free-space pseudopotential
approximation for the particle-particle interactions becomes
compromised \cite{blo02, tie00}.  In this case, one may obtain the
eigenenergies for the system by employing an energy-dependent
scattering length and solving for the energy eigenvalues
self-consistently \cite{bol02, bol03, bur02}.  For current purposes it
is supposed that we are in the regime for which the pseudopotential
approximation is still valid and the 1D collisional coupling,
$g_{1D}$, acts as a parameter for the system. This regime requires
that the range of the interparticle interaction be much smaller than
the characteristic length scale of the confining potential
\cite{blo02, tie00, bol02, bol03, bur02} (i.e.  $a_{1D} \ll d$).

\section{Single-particle eigenstates}
\label{sect:single_particle}

We can rewrite the single particle Hamiltonian in
eq.~\eqref{eq:SingleParticleHamiltonian} according to the rescaling $
x = d \bar{x}$, where $d$ is the ground state extent in the axial
direction, as introduced above,
\begin{equation}
 \label{eq:sin_part_ham_scaled}
 \bar{h} = -\frac{1}{2}\frac{\partial^2}{\partial 
   \bar{x}^2} + \frac{1}{2} \bar{x}^2 + \bar{\kappa} \delta 
 ( \bar{x})\;, 
\end{equation}  
where the scaled barrier strength is now given by $\bar{\kappa} = (
\hbar \omega d)^{-1} \kappa$.  The time-independent Schr\"odinger
equation for this system then reads
\begin{equation}
 \label{eq:tise_scaled}
 \bar{h}\phi_n(\bar{x}) = \bar{E}_{n} \phi_{n}(\bar{x})\;. 
\end{equation}    
Due to the scaling, the energies $\bar{E}_{n}$ are given in units of 
$\hbar\omega$.  At this point, for convenience, we drop the `bar' on 
all quantities, and acknowledge that we are, henceforth, dealing in 
the scaled quantities just described.  The analytic solution to
eq.~(\ref{eq:tise_scaled}), for those eigenfunctions of even symmetry,
can be found as, \cite{bus98},
\begin{equation}
 \label{eq:sin_part_sym_vec}
 \phi_n (x) = \mathcal{N}_n\; e^{-\frac{x^2}{2}} U \left(
 \frac{1}{4} - \frac{E_n}{2}, \frac{1}{2}, x^2 \right) \quad n =
 0, 2, 4 \ldots\;.
\end{equation}
Here $\mathcal{N}_n$ is the normalization constant and $U(a,b,z)$ are
the Kummer functions \cite{abr72}. The corresponding eigenenergies,
$E_n$, are determined by the roots of the implicit relation,
\cite{bus98}
\begin{equation}
 \label{eq:sin_part_sym_val}
 -\kappa = 2 \frac{\Gamma\left( -\frac{E_n}{2} + \frac{3}{4}\right)}
                  {\Gamma\left( -\frac{E_n}{2} + \frac{1}{4}\right)}\;.
\end{equation}
By contrast, the antisymmetric eigenfunctions vanish at the origin and
are unaffected by the barrier.  They are therefore given by the odd
eigenstates of the unperturbed harmonic potential ($\kappa =0$)
\begin{equation}
  \label{eq:sin_part_anti_vec}
 \phi_n(x) = \mathcal{N}_n H_n (x) e^{-\frac{x^2}{2}} \quad n =1,3,5\ldots\;,
\end{equation}
where $H_n(x)$ is the $n^{th}$ order Hermite polynomial.  The
corresponding energies are given by the eigenvalues of the odd parity
states of the harmonic oscillator, $E_n = \left( n + \frac{1}{2}
\right)$.

Considering eq.~\eqref{eq:sin_part_sym_val} in the limit $ \kappa
\rightarrow 0 $, we find $ E_n = \frac{1}{2}, \frac{5}{2},
\frac{9}{2},\ldots $, and the even eigenstates are simply given by the
even harmonic oscillator solutions.  On the other hand, for $ \kappa
\rightarrow \infty $, these energies converge towards $ E_n =
\frac{3}{2}, \frac{7}{2}, \frac{11}{2}, \ldots$, and each even
eigenstate becomes degenerate with the next highest-lying odd parity
state.

\section{Tonks Molecule}
\label{sect:two_bosons_tg}

In the limit $g_{1D}\rightarrow\infty$ the point-like, impenetrable, 
interaction between the two atoms can be represented as a constraint 
on the allowed bosonic wavefunction, $\Psi^B_k$, \cite{gir00, gir01, 
bus03, yuk05},
\begin{equation}
 \label{eq:hard_core_constraint}
 \Psi^B_k ( x_1, x_2 ) = 0 \quad\text{if}\quad x_1 = x_2\quad 
 \text{for all } k\;, 
\end{equation}
where $k$ is an index labelling the eigenstates. One can see
immediately that this constraint is equivalent to the exclusion
principle for a corresponding system of two spin-aligned fermions,
which is a symmetry that gives rise to the Bose-Fermi mapping theorem
\cite{gir60, gir00, yuk05}. It allows one to solve the strongly 
interacting system of two bosons by solving the, often more 
accessible, system of two non-interacting fermions, then properly 
symmetrising the final wavefunction. In particular, the ground state 
of the two-boson system (the Tonks molecule), $\Psi^B_0$, is related 
to the non-interacting fermionic ground state, $\Psi^F_0$, by
 \begin{equation}
 \label{eq:fermi_bose_mapping_gs}
 \Psi^B_0 (x_1,x_2) = | \Psi^F_0 (x_1,x_2)| \;.
 \end{equation}
The fermionic ground state, $\Psi^F_0 (x_1,x_2)$, is given by the 
Slater determinant of the two lowest single-particle orbitals, so
that,
 \begin{align}
 \label{eq:slater_determinant}
 \Psi^B_0(x_1,x_2)&=\frac{1}{\sqrt{2}}|\phi_0(x_1)\phi_1(x_2) 
                                     - \phi_0(x_2)\phi_1(x_1)|
				     \nonumber \\[0,2cm]
                  =&\frac{{\cal N}}{2}\;e^{-(x_1^2+x_1^2)} 
 \left|x_2 U \left( \frac{1}{4}-\frac{E_0}{2},\frac{1}{2},x_1^2\right)
             \right.\nonumber\\
      &\left.\qquad\qquad\quad
       -x_1 U \left( \frac{1}{4}-\frac{E_0}{2},\frac{1}{2},x_2^2\right)
 \right|\;,
 \end{align}
where $  {\cal N} $ is the normalization factor.

\begin{figure}[t]
   \includegraphics[width=\linewidth]{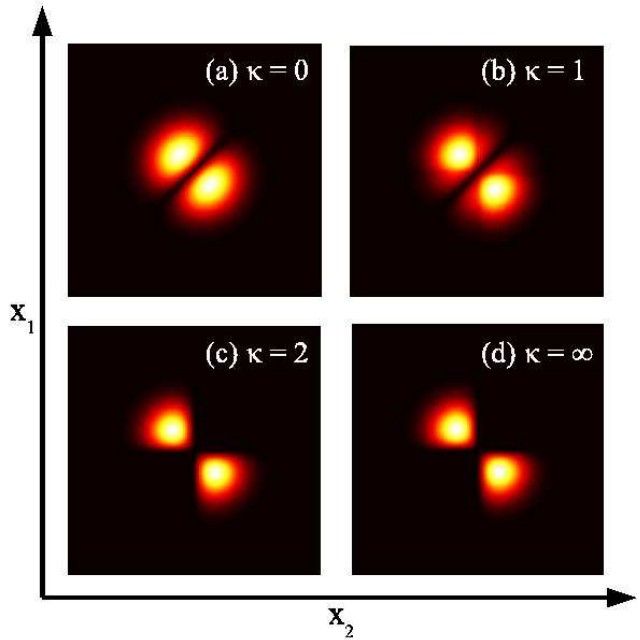}
   \caption{\label{fig:two_part_kappa} (Color online) Ground state
     wavefunction for a boson pair in a harmonic trap with a
     $\delta$-barrier along $x_1 = x_2 = 0$, of strength (a) $\kappa$ =
     0, (b) $\kappa = 1$, (c) $\kappa = 2$ and (d) $\kappa = \infty$.
     The corresponding, scaled ground state energies are $E_{0} = 2.0,
     2.4, 2.6$ and $3.0$, respectively. In each plot the horizontal
     and vertical axes run from -6 to +6, in scaled units.}
\end{figure}

Fig.~\ref{fig:two_part_kappa} shows the two-particle wavefunction for
the Tonks molecule in $(x_1,x_2)$ space, given by 
eq.~\eqref{eq:slater_determinant}, for different values of $\kappa$.
For $\kappa=0$ (Fig.~\ref{fig:two_part_kappa}(a)) the nodal line 
along $x_1=x_2$ reflects the infinite repulsion of the TG limit, or
equivalently the exclusion principle of
eq.~\eqref{eq:hard_core_constraint}.  The distribution of the
two-particle wavefunction shows a strong correlation between an 
$x_1>0$ and an $x_2<0$ coordinate, and vice versa.  Increasing 
$\kappa$ to 1, 2 and $\infty$ (Figs.~\ref{fig:two_part_kappa}(b),(c) 
and (d)), the wavefunction is reduced along the lines $x_1=0$ and
$x_2=0$ due to the strengthening potential barrier at the origin.  In
this process the wavefunction also becomes increasingly squeezed 
along the line $x_1 = -x_2$, indicating the localization of one 
particle on each side of the barrier.  We note that for values of 
$\kappa > 2$ there is no appreciable change in the two-particle 
density with barrier strength.

\subsection{Reduced single-particle density}
\label{subsect:rspd}

A quantity of fundamental importance in many-body physics is the
reduced single-particle density (RSPD), given by, \cite{col_re},
 \begin{equation}
 \label{eq:rspdm_def}
 \rho(x,x')=\int_{-\infty}^{+\infty} \Psi^B_0 (x,x_2)\Psi^B_0(x',x_2)dx_2\;.
 \end{equation}

\begin{figure}[t]
   \includegraphics[width=\linewidth]{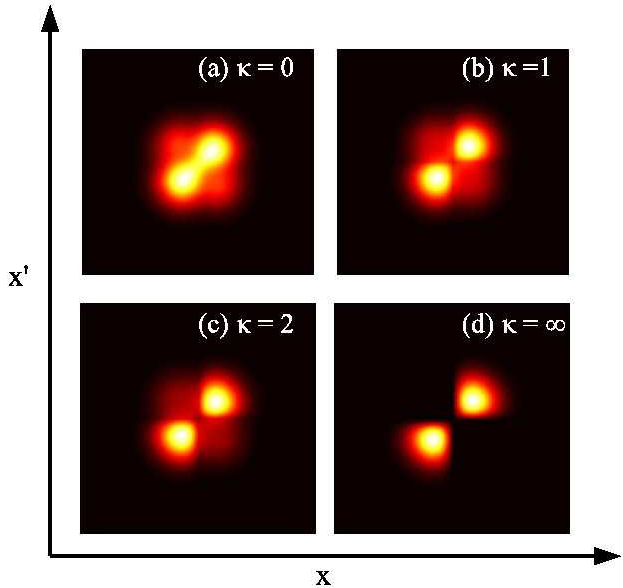} 
   \caption{\label{fig:rspdm_kappa} (Color online) Reduced 
     single-particle density matrix for the Tonks molecule, for
     barrier strength (a) $\kappa = 0$, (b) $\kappa = 1$ , (c) 
     $\kappa = 2$ and (d) $\kappa = \infty$.  Each plot spans the 
     range $ -6 < x, x' < 6$.}
\end{figure}

The RSPD is illustrated in Fig.~\ref{fig:rspdm_kappa} for four
different barrier strengths, corresponding to the same values examined
in Fig.~\ref{fig:two_part_kappa}, i.e.~$\kappa = 0$ (a), $\kappa = 1$
(b), $\kappa = 2$ (c) and $\kappa = \infty$ (d).  The RSPD expresses
the self correlation and one can view $\rho (x,x')$ as the probability
that, having detected the particle at position $x$, a second
measurement, immediately following the first, will find the particle
at the point $x'$.  Classically, $\rho(x,x')=\delta(x-x')$, and one
can see from Fig.~\ref{fig:rspdm_kappa} that a strong enhancement of
$\rho(x,x')$ exists along the line $x=x'$.  In the absence of any
barrier, Fig.~\ref{fig:rspdm_kappa}(a), the significant off-diagonal
contributions reflect the delocalization of an individual particle,
since there is a non-vanishing probability that the second
measurement, $x'$, may find the particle anywhere in the trap.
Increasing the barrier strength, as seen in
Figs.~\ref{fig:two_part_kappa}(b), (c) and (d), leads to the emergence
of a quadrant separation.  For a stronger barrier the contributions in
the off-diagonal quadrants diminish.  In particular, in
Fig.~\ref{fig:two_part_kappa}(d) these off-diagonal contributions to
$\rho(x,x')$ vanish altogether.  The strong barrier restricts tunnelling
from the left side of the well to the right, and vice versa.  In this
scenario, the ground state of the system is comprised of each member
of the boson pair in a separate half-well.

\subsection{ Momentum distribution}
\label{subsect:mom_dist_tg}

\begin{figure}[t]
   \includegraphics[width=\linewidth]{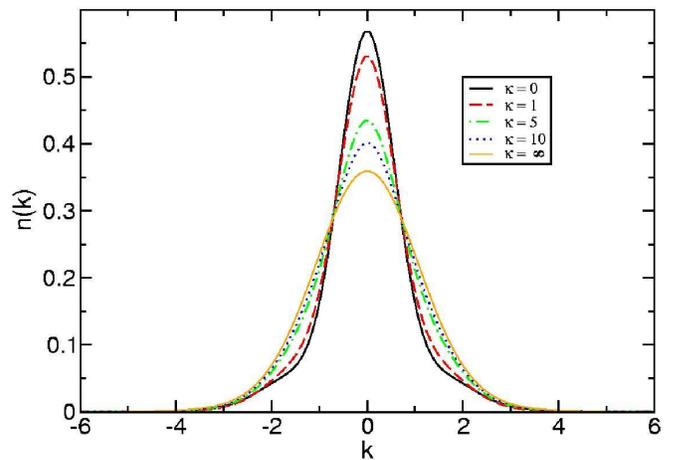}
   \caption{\label{fig:mom_dist_tg} (Color online) Momentum
     distribution, $n(k)$, for $\kappa=0,1,5$ and $10$, with the
     corresponding normalised ground state energies $ E_0= 2, 2.4,
     2.8$ and $2.9$. The momentum distributions broaden for increased
     barrier strength due to the localization of the particles in
     separate halves of the trap. Also shown is the momentum
     distribution for the $\kappa = \infty$ case, given by
     eq.~(\ref{eq:anal_mom_dist_kappa_inf_tg}), for which $E_0 =
     3.0$.}
\end{figure}

While, due to the Bose-Fermi mapping theorem, the density
distributions of a sample of bosons and fermions becomes identical in
the TG limit, the momentum distribution can still be used for
distinction \cite{gir01} . The reciprocal momentum
distribution, $n(k)$, is calculated from the reduced single-particle
density
\begin{equation}
 \label{eq:mom_dist_int}
 n(k) \equiv (2\pi)^{-1} \int_{- \infty}^{+ \infty}\int_{- \infty}^{+ \infty} 
 \rho (x,x')e^{- \imath k(x-x')} dx\; dx'\;,
\end{equation}
where $\, \int_{- \infty}^{+ \infty} n (k) dk = 1 \,$.  Equivalently,
one may obtain the momentum distribution for this system by
considering the diagonalization of $\rho(x,x')$. The eigenvalue
equation to be solved is
\begin{equation}
 \label{eq:rspdm_diag}
 \int_{-\infty}^{+\infty}\rho(x,x') \psi_i(x')dx' = \lambda_i\psi_i(x)\;,
\end{equation}
where the eigenvalue, $\lambda_i$, represents the fractional
population of the `natural orbital' $\psi_i(x)$ such that $\sum_i
\lambda_i = 1$.  Using a discretized form for the quadrature allows
one to rewrite the integral equation (\ref{eq:rspdm_diag}) as a linear
algebraic equation.  The momentum distribution, $n \left( k \right)$,
may then be obtained from the relation
\begin{equation}
 \label{eq:mom_dist}
 n(k) = \sum_i\lambda_i| \mu_i(k) |^2\;,
\end{equation}
where $\mu_i(k)$ denotes the Fourier transform of the natural orbital
$\psi_i(x)$,
\begin{equation}
  \label{eq:ft_nat_orb}
  \mu_i(k)=\frac{1}{\sqrt{2 \pi}}\int_{-\infty}^{+\infty} 
  \psi_i(x)e^{- \imath k x}dx\;.        
\end{equation}
Fig.~\ref{fig:mom_dist_tg} shows the momentum distribution in the TG
limit for four different values of the barrier strength, $\kappa = 0$,
1, 5 and 10.  As the barrier strength is increased the momentum
distribution becomes broader.  This observation is consistent with the
earlier observation that the two separate particles become
individually localized in the two separate half-wells.

In the limit of infinite barrier strength ($\kappa=\infty$) the system
becomes doubly degenerate which allows us to calculate an analytical
expression for the momentum distribution.  By changing the
computational basis and defining $\eta(x)=\frac{1}{\sqrt{2}} \left[
\phi_0(x) + \phi_1(x) \right]$, where $\phi_0$ and $\phi_1$ are the
ground and first-excited eigenfunctions of the single-particle 
Hamiltonian \eqref{eq:sin_part_ham_scaled}, with
$\phi_0(x)=|\phi_1(x)|$, we find that the wavefunction is only 
finite in the region $x>0$. The momentum distribution is the given by
the direct Fourier transform of $\eta(x)$
 \begin{equation}
 \label{eq:anal_mom_dist_kappa_inf_tg}
 n(k) = \frac{2}{\pi^\frac{3}{2}} \left\{ \left[ 1 - k^2
     e^{-\frac{k^2}{2}} M \left( \frac{1}{2}, \frac{3}{2}, \frac{1}{2}k^{2} 
     \right) \right]^{2} + \frac{\pi}{2} k^{2} \textrm{e}^{-k^{2}} \right\} 
 \;.
 \end{equation}
 This analytic momentum distribution, for the case $\kappa = \infty$,
 is also plotted in Fig.~\ref{fig:mom_dist_tg}.  It can be seen that,
 in the limit of large $\kappa$, the momentum distribution calculated
 from the diagonalisation of the RSPD matrix tends towards the profile
 given by eq.~\eqref{eq:anal_mom_dist_kappa_inf_tg}.
 
\subsection{\label{subsect:von_neumann_entropy_tg} Ground state entropy}

Entanglement is not only a fundamental quantity in quantum mechanics,
it is also one of the most important resources in quantum information
theory, where it is often responsible for the increased efficiency of
quantum algorithms over their classical counterparts.  Previous
authors have shown that the von Neumann entropy is a good measure of
entanglement for a system of two bosons \cite{pas01, li01, sun06}.  In
the case of indistinguishable particles , however, differentiating
between entangled and non-entangled states requires that one considers,
simultaneously, both the von Neumann entropy of the reduced
single-particle density and the \textit{Schmidt number} \cite{ghi03,
  ghiA04, ghiB04}.  The Schmidt number is given by the number of
non-zero eigenvalues, $\lambda_i$, of the reduced single-particle
density, $\rho$ (see eq.~(\ref{eq:rspdm_diag})).  In this work we
shall use the von Neumann entropy to quantify the entanglement in the
position coordinates, $x_1$ and $x_2$, of the boson pair and the
Schmidt number shall only be discussed when it affects the
interpretation of the results presented.

\begin{figure}[t]
   \includegraphics[width=\linewidth]{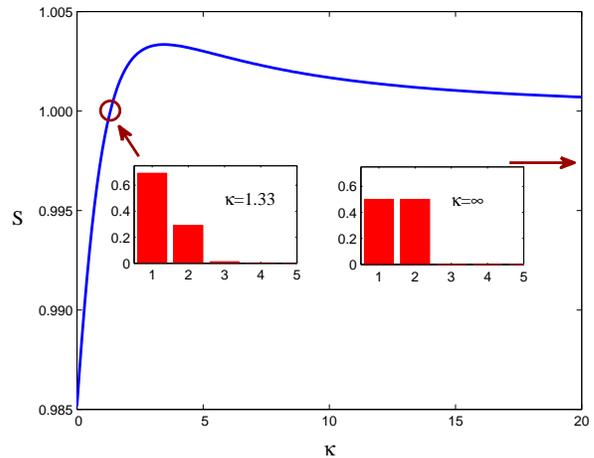}
   \caption{\label{fig:von_neumann_tg}(Color online) Von Neumann 
     entropy, $S$, for the Tonks molecule, as a function of the 
     barrier strength, $\kappa$.  When $\kappa = 0$ then $S \approx
     0.985$.  As the barrier is strengthened the entropy increases to
     a maximum at $\kappa \approx 3.4$.  In the limit of $\kappa 
     \rightarrow \infty$ then $S \rightarrow 1$, corresponding to a 
     non-entangled state.  The bar charts show the values of the 
     Schmidt numbers at the point where $( \kappa = 1.33, S = 1 )$ and 
     for the limit $( \kappa = \infty, S = 1)$.}
\end{figure}

The von Neumann entropy, $S$, is defined by
\begin{equation}
 \label{eq:von_neumann}
 S = -\sum_{i} \lambda_{i} \log_2 \lambda_{i} \;.
\end{equation}
and we calculate the values for $\lambda_{i}$ by numerically
diagonalising the RSPD matrix as a function of $\kappa$. The results
are shown in Fig.~\ref{fig:von_neumann_tg}. Interestingly, one sees
that the entropy begins at a value of about 0.985 for $\kappa = 0$,
which agrees well with the limiting value suggested in \cite{sun06}.
As $\kappa$ increases, $S$ is seen to increase through a value of
unity.  It peaks for $\kappa \approx 3.4$ (corresponding to a ground
state energy of $E_{0} \approx 2.85$) before dropping off and tending
towards unity in the limit $\kappa \rightarrow \infty$, corresponding
to a non-entangled state.  Identification of this state as 
non-entangled follows from the fact that the von Neumann entropy for
this state (with an infinite barrier) equals unity and the Schmidt
number is found to equal 2, \cite{ghi03,ghiA04,ghiB04}.  In this
situation, the ground state of the boson pair is comprised of one
particle residing in the left half-well and one in the right.
However, owing to the indistinguishability of the particles, one
cannot say which particle resides to the left and which to the right.
This lack of information, arising solely from the indistinguishability
of the particles, leads to the value of 1 for the von Neumann entropy.
Pure statistical correlations are of little intrinsic value to any
quantum information protocol, and the state is regarded as
non-entangled.

By contrast, the point at which $S=1$ for the finite value of $\kappa
\approx 1.33$, represents an entangled state.  This is due to the fact
that the Schmidt number at this point is $>2$, allowing one to
classify the state as truly entangled, beyond purely statistical
correlations, \cite{ghi03, ghiA04, ghiB04}.

\section{ Variable interaction strength}
\label{sect:two_particles_numerical}

For finite particle interactions no analytical solution to the
inhomogeneous two-particle problem is known (except in the case of
$\kappa=0$, \cite{bus98}).  In this section we, therefore, use a
numerical discretization scheme to study the ground state properties
of the boson dimer as a function of varying interaction strength, as
well as barrier strength.

Discretization of the spatial coordinates $x_{1}$ and $x_{2}$ is
achieved by means of a discrete variable representation (DVR),
\cite{bay86, lig00}.  The two-particle wavefunction is represented by 
the direct product
 \begin{equation}
 \label{eq:dvr_var_wavefunc}
 \Psi(x_1,x_2) =\sum_{i,j=1}^N \Psi_{ij} f_i(x_1) f_j(x_2)\;.
 \end{equation}
Here $\Psi_{ij}$ is the value of the two-particle wavefunction at the
mesh point $(x_{1}=q_{i},x_{2}=q_{j})$, with $i,j = 1,2,\ldots,N$.  
Clearly, these mesh points are finite in number and will be restricted
to some region in $(x_1,x_2)$ space, defined by the boundaries $a$ and
$b$, such that
 \begin{equation}
 \label{eq:mesh_boundary}
 a < q_{i} < b \hspace*{0.5cm} i = 1, 2, \ldots, N \;.
 \end{equation}
The values, $\Psi_{ij}$, play the role of variational parameters to be
found and the $f_i(q)$ are a set of $N$ Lagrange functions which have
the property that they are localized about the mesh points $q_{1},
q_{2}, \ldots, q_{N}$.  In addition to satisfying the usual 
interpolation conditions,
 \begin{equation}
 \label{eq:interpolation_condition}
 f_i(q_{j}) = \delta_{ij}\qquad \forall \; i, j\;,
 \end{equation}
one also requires that these Lagrange functions satisfy the 
orthogonality condition
 \begin{equation}
 \label{eq:orthogonality_condition}
 \int_a^b f^*_i(q)f_j(q) dq =\lambda_i \delta_{ij}\;.
 \end{equation}
Here $\lambda_{i}$ are the generalised Christoffel numbers associated
with the mesh, \cite{bay86}, and 
 \begin{equation}
 \label{eq:christoffel_numbers}
 \lambda_{i} = 1 \hspace*{0.5cm} \forall \; i \; ,
 \end{equation}
for the Cartesian mesh considered in this work.  For this Cartesian 
mesh the Lagrange functions are given by
 \begin{equation}
 \label{eq:cartesian_mesh}
 f_i(q)=\frac{1}{N}\frac{\sin[\pi(q-i)]}{\sin[\pi(q-i)/N]}\;.
 \end{equation}
Using the basis expansion of eq.~\eqref{eq:dvr_var_wavefunc} in the
Schr\"odinger equation \eqref{eq:tise_scaled} results in a discrete
eigenvalue problem that can be solved using standard linear algebra
techniques.

\subsection{ Reduced single-particle density}
\label{subsect:rspd_dvr}

\begin{figure}[t]
  \includegraphics[width=\linewidth]{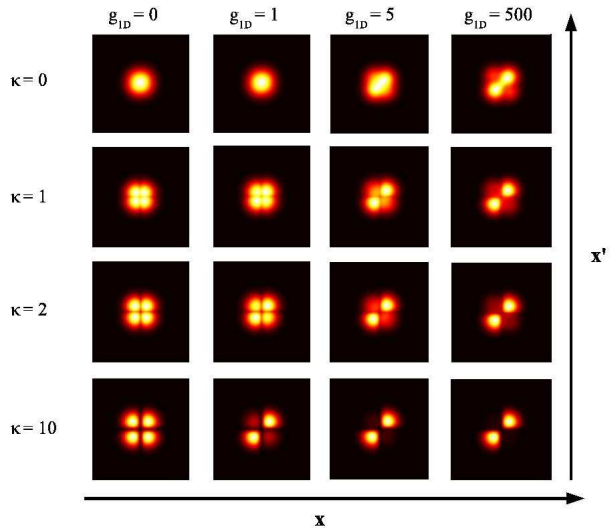}
  \caption{\label{fig:rspdm_dvr} (Color online)
    RSPD, $\rho(x,x')$ as a function of interaction strength $g_{1D} =
    0, 1, 5$ and $500$ and barrier strength $\kappa = 0, 1, 2$ and $10$.
    Each individual plot spans the range $-6.4 < x, x' < 6.4$.}
\end{figure}

We have calculated $\rho(x,x')$ using $N = 81$ mesh points in each
dimension (i.e.~$x_1$ and $x_2$) and a mesh spacing of $\Delta x =
0.16$.  Color density plots of the reduced single-particle density are
presented in Fig.~\ref{fig:rspdm_dvr} for four different values of 
interaction strength and four different values of barrier strength. 
Each row illustrates the transition from a non-interacting pair 
($g_{1D} = 0$) to a strongly interacting dimer ($g_{1D} = 500$), and 
each column illustrates the transition from a single well ($\kappa =
0$) to an, essentially, split trap ($\kappa = 10$).

In the first column of Fig.~\ref{fig:rspdm_dvr} the non-interacting
limit ($g_{1D} = 0$) is considered.  The increased barrier strength at
the origin manifests itself by diminishing $\rho (x,x')$ along the
lines $x=0$ and $x'= 0$, thus partitioning the structure into four
quadrants.  The even division of $\rho(x,x')$ over all four quadrants
reflects the delocalization of each individual particle over the two
half-wells.

The second column of Fig.~\ref{fig:rspdm_dvr} shows the same color
density plots for $\kappa = 0,1,2$ and 10 for a finite interactions
strength of $g_{1D} = 1$.  In the absence of a barrier ($\kappa = 0$)
the RSPD exhibits similar features to the non-interacting case,
although it expands slightly in both $x$ and $x'$. Strengthening the 
barrier again gives rise to a quadrant structure.  However, the 
presence of repulsion reduces $\rho(x,x')$ in the off-diagonal 
quadrants, meaning that the initial detection of a particle in the 
left half-well precludes its subsequent detection in the right 
half-well and vice versa.  Analogously to the Bose-Hubbard model, the 
system will be governed by the interplay between the tunnelling 
(determined by the strength of the barrier, $\kappa$) and the on-site 
interaction (determined by the interaction parameter, $g_{1D}$). For a
strong barrier ($\kappa = 10$) and finite interaction, there is a 
blockade and the insulator state dominates, with one boson in each 
half-well.  In terms of the reduced single-particle density, 
$\rho(x,x')$, this leads to the vanishing of the off-diagonal 
contributions as tunnelling of a given particle between the two 
half-wells becomes increasingly unlikely.  This behaviour is 
increasingly visible in the third and fourth column when the 
interaction strength is increased to $g_{1D} = 5$ and 500, 
respectively.

As the interaction strength increases, the first plot in the third
column shows a clear deviation from the circular structure observed in
the $\kappa=0$ case for lower interaction strength.  The distribution
is now clearly enhanced along the line $x = x'$ and reduced in the
direction orthogonal to this.  The stronger repulsive interaction has
the effect of reducing the `delocalization' of the particles.  As the
barrier strength is increased the off-diagonal contributions die-off
faster than in the case $g_{1D} = 1$.  This is due to the stronger
interactions encouraging the localization of the particles at even
smaller barrier strengths.  The superfluid character, that is 
indicated by the quadrant structure, decays already for smaller 
values of,$\kappa$. As in the case of $g_{1D}=1$, as the barrier 
strength is,increased one observes the reduction in the off-diagonal 
contributions and in the limit of large $\kappa$ one observes the, 
almost perfect, localization of the two particles in the two separate 
half-wells.

Finally, the last column illustrates $\rho(x,x')$ for very strong
repulsion, $g_{1D} = 500$.  As one expects, the reduced
single-particle densities closely resemble the plots displayed in
Fig.~\ref{fig:rspdm_kappa} for the Tonks molecule.

\subsection{ Momentum distribution}
\label{subsect:mom_dist_dvr}

\begin{figure}[t]
  \includegraphics[width=\linewidth]{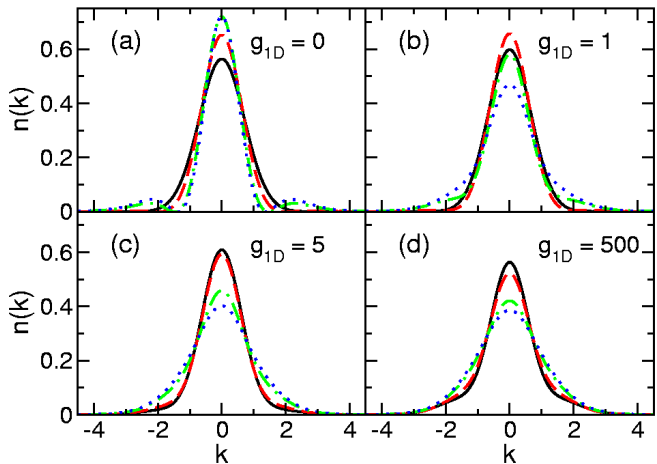}
  \caption{\label{fig:mom_dist_dvr} (Color online) Momentum
    distributions for varying interparticle interaction strength
    $g_{1D} = 0$ (a), 1 (b), 5 (c) and 500 (d).  Within each plot the
    distribution is considered for different values of the barrier
    strength: $\kappa$ = 0 (solid line, black), 1 (dashed line, red), 
    5 (dash-dot line, green), 10 (dotted line, blue).  These 
    calculations have been carried out by means of the DVR 
    discretization of the spatial coordinates $x_1$ and $x_2$, with 
    $N = 61$ DVR mesh points in each dimension and a scale factor of 
    $\Delta x = 0.16$.} 
\end{figure}

The momentum distributions, $n(k)$, can be obtained from the reduced
single-particle density, $\rho(x,x')$, using the same methods outlined
in Sec.~\ref{subsect:mom_dist_tg}.

Fig.~\ref{fig:mom_dist_dvr}(a) shows the distribution obtained for two
non-interacting particles ($g_{1D} = 0$) for varying $\kappa$. In this
case, the momentum distribution is given by the square of the
single-particle wavefunction in momentum space, $|\chi(k)|^2$.  In the
limit of an infinitely strong barrier the single-particle wavefunction
becomes $\phi_0(x) = (2/\sqrt{\pi})^{1/2}|x| e^{-x^{2}/2}$ and we can
calculate the momentum distribution analytically,
\begin{equation}
 \label{eq:anal_mom_dist_kappa_inf}
 n(k) = \frac{4}{\pi^{3/2}} \left[1-k^2e^{-\frac{k^2}{2}} 
        M\left(\frac{1}{2},\frac{3}{2},\frac{1}{2}k^2\right)\right]^2\;.
\end{equation}
The interplay between the first and second terms give rise to the
secondary peaks seen in Fig.~\ref{fig:mom_dist_dvr}(a) for $\kappa =
2,5$ and 10.  Physically, these peaks arise due to the interference of
the particle, split between the two separate half-wells, in analogy 
with a double-slit arrangement. When increasing the interaction 
strength (Figs.~\ref{fig:mom_dist_dvr}(b)-(d)) these secondary peaks 
disappear, which can be attributed to the increased localization of 
the particles.

At the same time as observing the emergence of these secondary peaks
with increased barrier strength for $g_{1D}=0$, one also observes a
narrowing of the central peak.  Strengthening of the barrier causes 
the ground state to shift upwards in energy, this shift will be 
accompanied by a spreading of the single-particle wavefunction in 
position space, which in turn gives rise to a reciprocal narrowing in 
momentum space. 

Figs.~\ref{fig:mom_dist_dvr}(c) and (d) display the momentum
distribution in the limit of strong repulsive interactions with the
same basic trends being observed in both plots.  Once again, as with
the reduced single-particle density, this fact suggests that the
behaviour of the ground state remains fairly constant for interaction
coupling $g_{1D} > 5$, such that these finite values of interaction
coupling will lead to behaviour which is qualitatively similar to the
regime of infinite repulsive interaction.  The results presented for
strong repulsive interaction ($g_{1D} = 500$) are expected to
correlate closely with the momentum distribution obtained in 
Sec.~\ref{subsect:mom_dist_tg} for the Tonks molecule, and a detailed
comparison of Figs.~\ref{fig:mom_dist_tg} and 
~\ref{fig:mom_dist_dvr}(d) verify that this is the case.

In the large interactions limit the momentum distribution for
$\kappa=0$ is observably different from the non-interacting case
(solid, black lines in Figs.~\ref{fig:mom_dist_dvr}(a) and (d)). In
particular non-Gaussian wings extending to higher $k$-value are
observed in the TG regime.  One may consider the trapping potential to
be switched off suddenly, and the two-particle wavefunction allowed to
expand freely.  In this case the wavefunction in coordinate space will
map on to that in momentum space, in the far field limit.  Clearly, 
the strong repulsion between the particles in the TG limit will lead 
to a proportion of the ensemble mutually recoiling at high speeds and 
in this way accounting for these high-$k$ wings in the momentum
distribution.  Increasing the strength of the barrier then has the
effect of broadening the momentum distribution as the individual
particles become localized to individual sides of the trap. The wings
in the momentum distribution are, therefore, a signal of a transition
into a Mott-insulator type state.  The spatial localization of the
particles is accompanied by a broadening in the momentum distribution
and this is the broadening observed in Figs.~\ref{fig:mom_dist_dvr}(c)
and (d).

\subsection{ Ground state entropy}
\label{subsect:von_neumann_dvr}

\subsubsection*{Variation of entropy with interaction strength}
\label{subsubsect:g1d_vary_kappa_fixed}

\begin{figure}[t]
  \begin{center}
    \includegraphics[width=\linewidth]{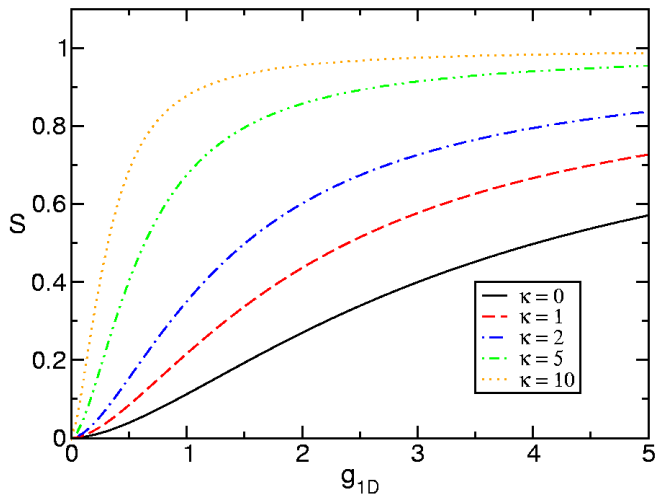}
    \caption{\label{fig:entropy_C_vary} (Color online) Effect of varying
      interaction strength ($g_{1D}$) on the von Neumann entropy ($S$)
      of the ground state.  The strength of the $\delta$-barrier is
      taken to be $\kappa = 0$ (thick solid line), 1 (dashed line), 2
      (dash-dot line), 5 (dash-dot-dot line) and 10 (dotted line). 
      It is seen that for increased strength of the central barrier,
      the entropy shows an increased sensitivity to the interaction 
      parameter about the value $g_{1D} = 0$.}
  \end{center}
\end{figure}

Let us first examine how the entropy of the two-particle system 
varies as one changes the interaction strength between the particles.
Similar calculations have been carried out by other authors,
\cite{sun06}, though restricted to a harmonic trap without a barrier.
This case is represented as the lowest (black) line in
Fig.~\ref{fig:entropy_C_vary}.  For $g_{1D} = 0$ entanglement is
absent and $S = 0$.  As the interaction strength is increased the 
entanglement increases.  For $g_{1D} \rightarrow \infty$ the 
entropy saturates at a value of $S \sim 0.985$ in the absence of any 
barrier, \cite{sun06}.  Also shown in Fig.~\ref{fig:entropy_C_vary} 
is the variation in the entropy with interaction strength when one 
introduces a $\delta$-barrier at the  well centre.  Four different 
barrier strengths are plotted: $\kappa = 1$ (dashed line), 2 
(dash-dot line), 5 (dash-dot-dot line) and 10 (dotted line).  One 
striking behaviour is noted, as one increases the barrier strength, 
the sensitivity of $S$ to small changes in $g_{1D}$, about $g_{1D} =
0$, is dramatically increased. As the barrier strength is increased 
the harmonic trap is split into two half-wells and the tunnelling 
between these two half-wells is made increasingly unfeasible.  As a 
consequence, the ground state of the two-particle system is less 
capable of adapting to changes in the interaction strength between 
the particles, leading to an increased sensitivity of the entropy in 
this respect.  For all values of $\kappa$, as $g_{1D} \rightarrow
\infty$ the entropy tends to a value close to unity (see
Fig.~\ref{fig:von_neumann_tg}).

\subsubsection*{Variation of entropy with barrier strength}
\label{subsubsect:kappa_vary_g1d_fixed}

\begin{figure}[t]
  \begin{center}
    \includegraphics[width=\linewidth]{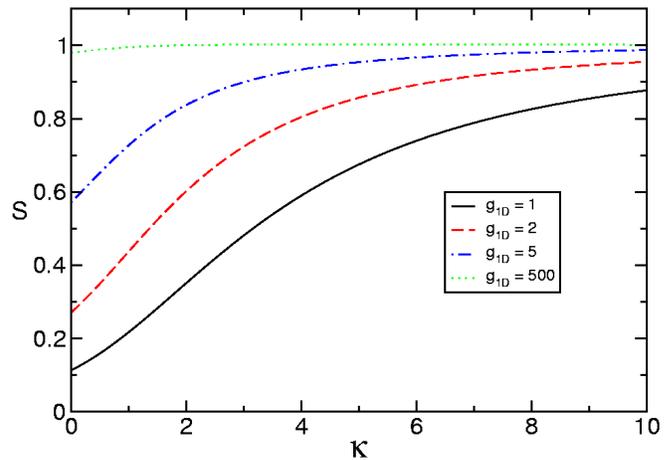}
    \caption{\label{fig:entropy_kappa_vary} (Color online) Effect of
      varying barrier strength on the von Neumann entropy, $S$, of the
      ground state.  The strength of the inter-particle interaction is
      set to $g_{1D} = 1$ (solid line), 2 (dashed line), 5 (dash-dot
      line) and 500 (dotted line).  The initial value of the entropy
      (i.e.~in the absence of any barrier) is dictated by the strength
      of the interparticle interaction, with larger interaction
      leading to increased entropy.  One observes that in all cases,
      in the limit of a strong barrier, the von Neumann entropy
      saturates at a value $S=1$.}
  \end{center}
\end{figure}
Finally, we present the results of how the entropy of the two-particle
system changes with the strength of the central barrier in
Fig.~\ref{fig:entropy_kappa_vary}.  For the case of zero interactions
the entropy remains zero for all barrier strengths.  In the presence 
of a finite interaction the entropy begins with a non-zero value,
representing the entanglement in the harmonic trap with no barrier.
As the barrier is strengthened the entropy increases gradually towards
unity, and saturates at this value.  As was discussed in
Sec. \ref{subsect:von_neumann_entropy_tg}, in the limit of infinite
barrier strength, the two-particle system will become non-entangled.
This is due to the fact that, in order to minimize the energy of the
system, the repulsively interacting particles will localize on
opposite sides of the well.  The only correlations that then exist
between the particles can be attributed to their indistinguishable
nature.

In the limit of vanishing barrier strength, the larger the interaction
strength, the larger is the initial value of the entropy.  As a
consequence, for larger values of $g_{1D}$ the entropy changes less
dramatically as the barrier strength is increased.  It is noted that for
the case of $g_{1D} = 500$ (dotted line) we are effectively
considering the TG regime.  From Fig.~\ref{fig:entropy_kappa_vary} it
appears that in this regime the entropy remains close to unity for all
values of $\kappa$.  However, closer inspection of these numerical
results reveals that this curve actually follows the same trend as
illustrated in Fig.~\ref{fig:von_neumann_tg}, obtained from the
analytical treatment of the Tonks molecule.  This further illustrates
the correspondence of these DVR mesh calculations to the analytical 
TG results in the limit of infinite repulsive interactions.

\section{ Conclusions}
\label{sect:conclusions}

In the present work we have carried out a detailed examination of the
ground state for two particles in a $\delta$-split harmonic trap. We
have found that in the presence of interactions the reduced
single-particle density exhibits vanishing contributions in the
off-diagonal quadrants in the limit of increasing barrier strength.
This feature is attributed to the localization of individual particles
on either side of the split trap, a situation analogous to the Mott
insulator regime in lattice studies and also reflected in the
corresponding momentum distributions.  More specifically, in the
non-interacting case with a strong barrier one observes secondary peaks
in the momentum density, attributed to interference.  These secondary
peaks vanish in the presence of interaction owing to the localization
of individual particles.  In the Tonks-Girardeau limit, increasing the
barrier strength has the effect of broadening the momentum distribution,
a feature that may be explained in terms of the squeezing of the
wavefunction for the system in position space.  Finally, we have shown
that the von Neumann entropy for this system is sensitive to the two
parameters of interaction strength and barrier strength.  For a given
barrier strength, an increasingly repulsive interaction strength will
cause the von Neumann entropy to saturate at a value close to unity.
It is found that increasing the strength of the barrier has the effect
of making the von Neumann entropy increasingly sensitive to small 
changes in the interaction coupling about the value of zero coupling. 
At the same time, for a fixed value of interaction strength, 
increasing the barrier strength has the effect of increasing the 
entropy of the system.  In the limit $\kappa \rightarrow \infty$ the 
entropy saturates at a value of unity.

We would like to remark that even though our analysis makes use of an
idealised $\delta$-function potential, such an approximation is known
to, not only, encapsulate the basic physics, but can also be a very 
good approximation to experimental setups. These include wide traps 
that are pierced by a highly-focused laser beam as well as, for
example, the situation where a single particle of a different species 
is confined in the centre of the trap. Due to the low temperatures, 
the interaction with such a `quantum dot' would be well described by a
point-like potential.

\begin{acknowledgements}
  DSM and JFM thank Dr. M. Paternostro for valuable discussion.  DSM
  would also like to acknowledge funding from the Department for
  Employment and Learning (NI) and the support of the Sorella Trust
  (NI).  JG and TB would like to thank Science Foundation Ireland for
  support under project number 05/IN/I852.
\end{acknowledgements}

\bibliography{two_part_tg_split_trap}

\end{document}